\documentstyle[12pt] {article}
\begin {document}
\parindent=15pt
\begin{center}
\vskip 1.5 truecm
{\bf NEW APPROACH TO INDUCED QCD}\\
\vspace{1cm}
A.Shuvaev\\
Theory Department, St.Petersburg Nuclear Physics Institute\\
188350, Gatchina, St.Petersburg, Russia.
\end{center}
\begin{abstract}
Matrix model approach to multicolor induced QCD based on the quenched
momentum prescription is presented. It is shown that this model
exhibits the reduction of spatial degrees of freedom: the partition
function is
determined by the solution of one dimensional quantum mechanical
problem while the D-dimensional scalar field correlators coinside
with the same type correlators in the two-dimensional induced QCD.
\end{abstract}
\vspace{1cm}
{\bf 1.} Investigation of the large number of colors $N_c$ limit of QCD
permits
a deeper insight into the nature of the strong interaction. Analytical
summation of planar diagrams leads to a kind of masterfield equation. The
considerably reduction of actual degrees of freedom occurring
when $N_c\to\infty$ underlies this equation. The large number of
"angular-type" variables (unitary transformations) is effectively
"integrated out" in matrix models. The remaining variables form a meanfield
described in terms of the masterfield equation.

The case of induced QCD will be considered here. This model was introduced
by Kazakov and Migdal four years ago \cite{KMM}. Here the induced QCD is
revised but instead of the lattice version of QCD in Kazakov-Migdal model
(KMM) the prescription proposed by Gross and Kitazawa \cite{GK} for the
planar Feynman diagrams is adopted. As in the KMM the absence of the gluon
kinetic energy enables to integrate over the unitary matrix in closed form.
However contrary to the KMM case the explicit expression for the "angular"
integrals are not used in the present approach since it entirely avoids
the introduction of mean field as well as the solution of the masterfield
equation. The main goal of it is to establish the connection between 4D
induced QCD and the same lower dimensional theory.

\noindent
{\bf 2.} Since the Gross and Kitazawa prescription plays the crucial role
in the
following treatment it is instructive to begin with brief reminding of it.

The quantum scalar field in adjoint representation $\varphi(x)$ is replaced
by the $N_c\times N_c$ matrix $\varphi$ which does not depend on the $x_\mu$
variables. The field derivative is replaced by the commutator
$$
\partial_{\mu}\varphi \to i\,\bigl[\,P_{\mu}\,,\,\varphi\,\bigr],
$$
where $P_{\mu}$ are diagonal $N_c\times N_c$ matrixes:
\begin{equation}
\label{Pmu}
\bigl(P_{\mu}\bigr)^{ij}\,=\,\delta^{ij}\,p_{\mu}^i.
\end{equation}
The functional (euclidean) integral
\begin{equation}
\label{zf}
Z\,=\,\int \prod_x\,D\varphi(x)\,
\exp \left\{\int d^Dx\,Tr\left(-\frac 12\,(\partial_{\mu}\varphi)^2
-\frac 12 M^2\varphi^2-V(\varphi)\right)\right\}
\end{equation}
turns into the matrix integral
\begin{equation}
\label{zm}
Z\,=\,\int d\varphi\,\exp\left(\frac{2\pi}{\Lambda}\right)^D \left\{\,
Tr\left(\frac 12\bigl[P_{\mu},\,\varphi\bigr]^2
-\frac 12 M^2\,\varphi^2-V(\varphi)\right)\right\}
\end{equation}
where
$$
d\varphi\,=\,\prod_{ij}d\varphi_{ij}.
$$
When $N_c\to\infty$ the diagrams generated by the perturbative expansion
of the second integral coincide with the planar Feynman diagrams for the
first one, $\Lambda$ being the ultraviolet cutoff. Indeed the matrix model
propagator is
\begin{equation}
\label{matprop}
\langle\varphi_{ij}\,\varphi_{sp}\rangle\,=\,
\frac{\delta_{ip}\,\delta_{js}}{\bigl(p_{\mu}^i-p_{\mu}^j\bigr)^2+M^2}.
\end{equation}
Drawing the Feynman propogator by the doubleline one can assign to each
line its own momentum $p_1$ or $p_2$ so that the total momentum flowing
through the propagator is $k=p_1-p_2$. After this substitution the loop
integrals turn into the integrals over the momenta $p_{1,2}$. For
planar graphs the momentum loops are identical to the color ones inside
which the indexes $i,j$ in (\ref{matprop}) circulate. If the
D-dimensional hypercube in the momentum space is divided into $N_c$ equal
cells with the volume $\Lambda^D/N_c$ and the vectors $p_{\mu}^i$
in (\ref{matprop}) are choosen to lay inside these cells
the sums over the color indexes in
the matrix diagrams for the model (\ref{zm}) yield
the integral sums for the momentum loop integrals in the planar Feynman
diagrams for the functional integral (\ref{zf}). When $N_c\to\infty$
and $\Lambda$ is fixed the sums turn into the integrals. After this the
limit $\Lambda\to\infty$ can be taken. The hermitean $N_c\times N_c$
matrix $P_\mu$ (\ref{Pmu}) plays the role of the quenched momentum and
the components $p_{\mu}^i$ are the values it takes. \footnote{It is
really not necessary to identify the number of the terms in the integral
momentum sum with $N_c$. Indeed, while a planar Feynman diagram contribution
is $N_c^2\,G^F$ where $G^F$ does not depend on $N_c$, the rank $N$ matrix
model gives for a planar graph $G_{planar}\,=\,N^2\,\overline{G}$. The
quenched momentum prescription ensures $\overline{G}\,\to\,G^F$ for
$N\to\infty$ and it is this property that makes the matrix model to be
equivalent to the field theory. Thus only the planarity is important here.}

The unitary matrixes $e^{iPx}$ can be treated as a finite
dimension approximation to the space shift operator. One can define
$$
\varphi(x)\,\equiv \,e^{iPx}\varphi e^{-iPx}.
$$
Then the correlator
$$
\langle\,Tr\,\varphi(x_1)\,\cdots\,\varphi(x_n)\,\rangle
$$
calculated in the matrix model (\ref{zm}) in the leading $N_c$ order
will be the same as in the scalar field theory (\ref{zf}).

The incorporation of gauge field in this prescription is straightforward:
\begin{eqnarray}
A_\mu(x)\,&\to&\,A_\mu\nonumber\\
\frac 1i D_\mu\,&\to&\,P_\mu\,+\,\frac g{\sqrt{N_c}}A_\mu\nonumber\\
i\frac g{\sqrt{N_c}}G_{\mu\nu}\,&=&\,\bigl[\,D_\mu\,,\,D_\nu\,\bigr]
\nonumber
\end{eqnarray}
Here $A_\mu$ is a $N_c\times N_c$ hermitean matrix, $D_\mu$ is a
covariant derivative and the commutator in the last formula is understood
in the matrix sense. The (euclidean) matrix action can be written as
$$
S\,=\,-\frac 14 \left(\frac{2\pi}{\Lambda}\right)^D\,Tr \,G_{\mu\nu}^2\,+\,
S_{\rm gf}
$$
where $S_{\rm gf}$ is an appropriate gauge-fixing term.
The matrix transformation
$$
\frac g{\sqrt{N_c}}\,A_\mu\,\to\,V^{-1}A_\mu V\,+\,
\frac g{\sqrt{N_c}}\,V^{-1}\bigl[P_\mu,V\bigr]
$$
where $V$ is a unitary matrix
is equivalent through the relations $A_\mu(x)=e^{iPx} A_\mu e^{-iPx}$,
$V(x)=e^{iPx} V e^{-iPx}$ to a local gauge transformation in the field
theory.

Very important for the following is the additional constraint imposed on
the measure of the matrix integral, namely, the integration is carried out
only over the matrixes $A_\mu$ for which the covariant derivative $D_\mu$
has the same eigenvalues as the matrix $P_\mu$. An equivalent form of this
restriction is to rewrite the functional integral as
$$
Z\,=\,\int \prod_\mu dA_\mu c_\mu\,e^{-S}
$$
where
\begin{equation}
\label{constr}
c_\mu\,\sim\,\int dV_\mu\,\delta\bigl(\frac 1i D_\mu\,-
\,V_\mu^{-1}P_\mu V_\mu\bigr)
\end{equation}
and $dV_\mu$ denotes the invariant measure on the $SU(N_c)$ group. If it
were not for the constraint (\ref{constr}) the $P_\mu$ matrixes would be
completely excluded from the integral by shifting matrix variables
$\frac g{\sqrt{N_c}}A_\mu\to P_\mu\,+\,\frac g{\sqrt{N_c}}A_\mu=
\frac 1i D_\mu$.
The constraints (\ref{constr}) have a rather obvious meaning: any component
$A_\mu$ of the gluon field can be set to zero by the gauge transformation
$V_\mu$ special for each $\mu$ (only for the pure gauge field there is a
matrix $V$ common to all $\mu$).

\noindent
{\bf 3.} The induced QCD lagrangian describes the theory containing the
gluons and the adjoint scalars
\begin{equation}
\label{induced}
{\cal L}\,=\,\frac 12 Tr\bigl(D_\mu \varphi\bigr)^2\,-\,\frac 12 Tr
M^2\varphi^2
\end{equation}
but without the pure gluon term. The integration over the scalar field
yeilds the functional determinant
$$
Det \bigl(M^2-D^2\bigr)^{-\frac 12}\,=\,\exp\left\{-
\frac 1{96\pi^2}g^2 N_c\log\frac{\Lambda}{M}\,
\int d^4x\, Tr G_{\mu\nu}^2+O\bigl(1/M\bigr)\right\}.
$$
Here $O\bigl(1/M\bigr)$ denotes the terms finite when the ultraviolet cutoff
$\Lambda\to\infty$. They are suppressed by the powers of the scalar mass
$M$ which is assumed to be very large (although $M\ll\Lambda$). The theory
(\ref{induced}) seems to be similar in the limit
$M\to\infty (M\ll\Lambda)$ to the
gluodynamic, $g^2\sim 1/\log\Lambda/M$ being the coupling constant.
One can consider a more general case by adding to the lagrangian
(\ref{induced}) the scalar field interaction $V(\varphi)\,=\,\sum v_n
\varphi^n$.

According to the quenched momentum prescription the matrix model integral
for this theory in the hamiltonian gauge $A_0=0$ reads
\begin{eqnarray}
Z\,&=&\,\int\prod_\mu dA_\mu \,c_\mu\,\delta\bigl(A_0\bigr)\,d\varphi\,
\nonumber \\
&\times &\exp\left(\frac{2\pi}{\Lambda}\right)^D Tr\left\{\,
\frac 12\sum_\mu \bigl[D_{\mu},\,\varphi\bigr]^2
-\frac 12 M^2\,\varphi^2-V(\varphi)\right\}.\nonumber
\end{eqnarray}
Resolving $\delta$-functions in the constraints $c_\mu$ gives
\begin{eqnarray}
Z\,&=&\,\int\prod_\mu dV_\mu \,d\varphi\,\delta \left(\frac{N_c}g
\left(P_0\,-\,V_0^{-1}P_0 V_0 \right)\right)\nonumber \\
&\times &\exp \left\{\left(\frac{2\pi}{\Lambda}\right)^D\,
Tr\left(\frac 12\sum_\mu \bigl[V_{\mu}^{-1}P_\mu V_{\mu}\,,\varphi\bigr]^2
-\frac 12 M^2\,\varphi^2-V(\varphi)\right)\right\}. \nonumber
\end{eqnarray}
Writing here the scalar field matrix as $\varphi=U^{-1}\varphi_d U$
where $U$ is a unitary matrix and $\varphi_d$ is a diagonal one the
integral takes a form
\begin{eqnarray}
Z\,&=&\,\int\prod_\mu dV_\mu \,dU \,d\varphi_d \,
\Delta^2\bigl(\varphi_d\bigr)\,
\delta \left(\frac{N_c}g
\left(P_0\,-\,V_0^{-1}P_0 V_0 \right)\right)\nonumber \\
&\times &\exp\left(\frac{2\pi}{\Lambda}\right)^D \left\{\,
\frac 12\sum_\mu Tr \bigl[P_\mu\,,V_{\mu} U^{-1}\varphi_d U
V_{\mu}^{-1}]^2 \right.\nonumber \\
& & \left.-\frac 12 M^2\,\varphi_d^2-V(\varphi_d)\right\},
\nonumber
\end{eqnarray}
$V(\varphi_d)\equiv \sum_i V(\varphi_i)$ is the sum over the eigenvalues,
$\Delta(\varphi_d)$ is the Van der Monde determinant. Changing the
integration variables
\begin{eqnarray}
\label{change}
V_\mu \,U^{-1}\,&=&\,\widetilde{V}_\mu, \qquad \mu=1,\ldots,D-1,\\
V_0 \,U^{-1}\,&=&\,\widetilde{U}^{-1} \nonumber
\end{eqnarray}
allows to separate out the gauge-fixing term:
\begin{eqnarray}
Z\,&=&\,\int dV_0 \,\delta \left(\frac{N_c}g
\left(P_0\,-\,V_0^{-1}P_0 V_0 \right)\right)\,\cdot\,\int d\widetilde{U}
\prod_{\mu=1}^{D-1} d\widetilde{V}_\mu d\varphi_d \Delta^2\bigl(\varphi_d\bigr)
\nonumber \\
&\times &\exp \left(\frac{2\pi}{\Lambda}\right)^D\,\left\{
\frac 12 Tr\bigl[P_0\,,\widetilde{U}^{-1}\varphi_d \widetilde{U}]^2 +
\frac 12 Tr\sum_{\mu=1}^{D-1} \bigl[P_\mu\,,\widetilde{V}_{\mu}\varphi_d
\widetilde{V}_{\mu}^{-1}]^2 \right.\nonumber \\
& & \left.-\frac 12 M^2\,\varphi_d^2-V(\varphi_d)\right\}. \nonumber
\end{eqnarray}
Dropping the tildes and redefining $U_0=V_0$ the partition function can
be rewritten as ($a\sim 1/\Delta^2(P_0)$ is the normalization constant)
\begin{eqnarray}
Z\,=\,&a&\int d\varphi_d\, \Delta^2\bigl(\varphi_d\bigr)
\exp \left(\frac{2\pi}{\Lambda}\right)^D\,\left\{
-\frac 12 M^2\,\varphi_d^2-V(\varphi_d)\right\} \nonumber \\
\label{product}
&\times &\int \prod_\mu^D d V_\mu\,\exp \left(\frac{2\pi}{\Lambda}\right)^D\,
\left\{
\frac 12 \sum_{\mu=0}^{D-1} Tr\bigl[P_\mu\,,V_\mu\varphi_d V_\mu^{-1}]^2
\right\}.
\end{eqnarray}
The integration over the unitary matrixes in (\ref{product}) decays into
the product of $D$ equal integrals (they are equal since the matrixes
$P_\mu$ are unitary equivalent).

Consider the integral
\begin{equation}
\label{zphi}
Z_\varphi\,=\,\int dV\,\exp \left(\frac{2\pi}{\Lambda}\right)^D\left\{
\frac 12 Tr\bigl[P_\mu\,,V\varphi_d V^{-1}]^2 \right\}
\end{equation}
for the fixed matrix $\varphi_d$ and index $\mu$.
An explicit calculation of this expression is a rather nontrivial problem
although an essential progress has been made recently in studying the
similar type integrals \cite{Kaz}. Instead calculating it the much more
weak "scaling" property of this integral will be enough for the following.

The main contribution to the integral over the matrixes $\varphi$ in
(\ref{zphi}) comes from the domain where $\varphi_{ij}\sim 1$
and where the typical scale
of the eigenvalues $\varphi_d\sim N_c^{1/2}$. It is just the region that
gives the leading contribution of the order $N_c^2$ to the free energy
while the measure of the rest integration domain tends to zero in the
$N_c\to\infty$ limit. In this limit the eigenvalues distribution is
described by the smooth density function $\rho(\lambda)$.

Since $Tr\bigl[P_{\mu},\,\varphi\bigr]^2\sim N_c^2$ one can expect
\begin{equation}
\label{F}
Z_{\varphi}\,=\,\exp\left\{N_c^2F+O(N_c)\right\}
\end{equation}
where the coefficient $F$ depends on the eigenvalues of the matrix $\varphi$
and for the large $N_c$ can be treated as a functional $F=F[\rho]$.
Eq.(\ref{F}) is a consequence of the classical nature of the large $N_c$
limit which manifests itself in the factorization of the correlators of
colorless ($U(N_c)$ invariant) operators \cite{doug}.

The large $N_c$ behavior of the unitary matrix integral (\ref{zphi})
looks like as if it is dominated by a saddle point.
There is a representation which
makes the $N_c^2$ dependence in the integrand (\ref{zphi}) to be explicit
\cite{florat}. In the large $N_c$ limit $SU(N_c)$ algebra is equivalent to
the infinite dimensional Lie algebra of area preserving diffeomorphisms
of the sphere $SDiff(S^2)$. The matrix $A$ from $SU(N_c)$ algebra transforms
in this representation into the function $A(x_1,x_2,x_3)$ on the unit
sphere $x_1^2+x_2^2+x_3^2=1$, the commutator being replaced by Poisson
braket
$$
\lim_{N_c\to\infty}\frac {N_c}i\bigl[\,A\,,\,B\,\bigr]\,=
\,\bigl\{\,A\,,\,B\,\bigr\}
$$
which are defined as
$$
\bigl\{\,A\,,\,B\,\bigr\}\,=\,x_i\,\varepsilon_{ikl}\,\frac{\partial A}
{\partial x_k}\,\frac{\partial B}{\partial x_l}.
$$
The integral (\ref{zphi}) turns into the functional (infinite order) integral
over the group of area preserving diffeomorphisms
\begin{equation}
\label{zcont}
Z_{\varphi}\,=\,\int{\cal D}V\exp\left[\frac 12 \,N_c^2\,\left(\frac{2\pi}
{\Lambda}\right)^D\,\int d\Omega \bigl\{\,p\,,\,V(\varphi)\,\bigr\}^2
\right]
\end{equation}
where $V(\varphi)$ denotes the action of the diffeomorphism on the function
$\varphi$ and $\int d\Omega$ is the integral over the unit sphere. The
functions $\varphi$ and $p$ correspond to the matrixes $\varphi$ and $P_\mu$
in the integral (\ref{zphi}). Their particular structure as well as the
result of the action of $V$ on $\varphi$ is not important here. Only one
thing is significant for the following, namely, comparing the
expression (\ref{zcont}) with (\ref{F}) one can conclude that
$$
\int dV\,\exp\left\{-\frac 12\,\alpha\,\left(\frac{2\pi}{\Lambda}\right)^D\,
Tr\bigl[\,P\,,\,V^{-1}\varphi_D V\,\bigr]^2\right\}\,=\,
\exp\left\{\alpha \,N_c^2F+O(N_c)\right\}
$$
since it is nothing more than rescaling of $N_c$ by the factor
$\sqrt{\alpha}$. Thus the power of the integral $Z_{\varphi}$ can be replaced
within the leading order accuracy by the single integral
\begin{equation}
\label{scale}
\bigl(Z_{\varphi}\bigr)^D\,=\,\int dV\,\exp\left\{-\frac 12 \,D\,
\left(\frac{2\pi}{\Lambda}\right)^D\,Tr
\bigl[\,P\,,\,V^{-1}\varphi_d V\,\bigr]^2\right\}.
\end{equation}
The "scaling property" (\ref{scale}) will be the central point for the
following treatment. All other results are more or less trivial consequences
of it.

The relation (\ref{scale}) applied to the integral (\ref{product})
gives
\begin{eqnarray}
Z\,&=&\,a\int d\varphi_d \Delta^2\bigl(\varphi_d\bigr)\,dV\,
\nonumber \\
&\times &\,\exp \left(\frac{2\pi}{\Lambda}\right)^D\,Tr\,\left\{
\frac 12\,D\,\bigl[P_\mu\,,V\varphi_d V^{-1}]^2
-\frac 12 M^2\,\varphi_d^2-V(\varphi_d)\right\}.\nonumber
\end{eqnarray}
After combining the factors in the integration measure the partition function
(\ref{product}) takes the final form
\begin{equation}
\label{Z}
Z\,=\,a\int d\varphi\,\exp \left(\frac{2\pi}{\Lambda}\right)^D\,Tr\,
\left\{\frac 12\,D\,\bigl[P_1\,,\varphi_d]^2
-\frac 12 M^2\,\varphi_d^2-V(\varphi_d)\right\}.
\end{equation}
Here the matrix $P_\mu$ index is fixed by the value $\mu=1$, the result being
clearly independent on the particular choice of the space direction.

The expression (\ref{Z}) is a Gauss integral for $V(\varphi)=0$.
In the leading $N_c$ order
$$
\log Z\,=\,const\,+\,D N_c^2\,\log\Lambda\,-\,\frac 12\sum_{ik}\log
\bigl[D\bigl(p_i^1-p_k^1\bigr)^2\,+\,M^2\bigr]\,=
$$
$$
=\,const\,+\,D N_c^2\,\log\Lambda\,-\,\frac 12 N_c^2\,V_D
\left(\frac{\Lambda}{2\pi}\right)^{D-1}\int_{-\Lambda/2}^{\Lambda/2}
\frac{dp}{2\pi}\,\log\bigl(Dp^2+M^2\bigr)
$$
where $V_D$ is a total space volume\footnote{The value $V_{\rm tot}=(2\pi)^D
N_c/\Lambda^D$ is to be taken as a space volume in the quenched
model prescription to recover the planar graphs. Taking into account the
proper order of limits (first $N_c\to\infty$ and then $\Lambda\to\infty$)
one can put $V_{\rm tot}=N_c V_D$}. Collecting the terms independent of $M$
in the factor $\varepsilon_0$ one gets for the vacuum energy density
\begin{equation}
\label{fe}
\varepsilon\,=\,\varepsilon_0\,+\,\left(\frac{\Lambda}{2\pi}\right)^D\,
\left\{\log\left(1+\frac{M^2}{D\Lambda^2}\right)\,+\,\frac 2{\sqrt D}
\frac M\Lambda\arctan\sqrt D\frac M\Lambda\right\}.
\end{equation}

\noindent
{\bf 4.} When the scalar interaction $V(\varphi)\ne 0$ the evaluation of the
integral (\ref{Z}) is equivalent to solving the quantum mechanical problem.
Indeed, as is seen from inverting the quenched momentum prescription,
it coincides in the large $N_c$ limit with the partition function
\begin{equation}
\label{ZD}
Z_D\,=\,\int \prod_x {\cal D}\varphi(x)\,\exp Tr \int d^D x
\left\{-\frac 12\,D\,(\partial_1 \varphi)^2 - \frac 12\,M^2\,\varphi^2 -
V(\varphi)\right\}
\end{equation}
Since there is only one actual degree of freedom contributing to (\ref{ZD})
$$
Z_D\,=\,\exp \left\{-V_D\,\left(\frac{\Lambda}{2\pi}\right)^{D-1}\,E
\right\}
$$
where $E$ is the ground state energy for the one-dimensional system
described by the integral
\begin{equation}
\label{Z1}
Z_1\,=\,\int \prod_t {\cal D} Q(t)\,\exp N_c\,Tr \int d t
\left\{-\frac 12\, D \,\mu \,\dot{Q}^2 - \frac 12 \,\mu M^2 \,Q^2 -
V(Q)\right\}
\end{equation}
in which $\mu=\Lambda/2\pi$ and $V(q)=\sum_{n \ge 3} \mu^{1 + D(n/2-1)}
v_n q^n$. This relation immediately follows from the lattice version of the
theory (\ref{ZD}) which is equivalent to the ultraviolet regularization with
the cutoff $\Lambda$, $1/\mu$ being the lattice spacing.

The system (\ref{Z1}) is solved by reduction to the free fermions moving
in the external potential $V(q)$ \cite{Gross}. The energy is
$E=\sum_{i=1}^{N_c}{\cal E}_i$, where ${\cal E}_i$ are the energies of
the lowest $N_c$ occupied states:
$$
\left[-\frac 1{2 \mu D}\, \frac{\partial^2}{\partial x_i^2} + \frac 12\,
\mu M^2 \,x^2 + V(x)\right]\,\psi_i(x)\,=\,{\cal E}_i\psi_i(x).
$$
In the quasiclassical approximation which validity is justified by large
$N_c$
\begin{equation}
\label{feh}
E\,=\,N_c^2\,\int\frac{dp\,dq}{2\pi}\,H(p,q)\,\theta\bigl({\cal E}-H(p,q)
\bigr)
\end{equation}
where $\theta$ is the step function,
$$
H(p,q)\,=\,\frac{p^2}{2D}\,+\,\frac12 M^2q^2\,+\,\sum_{n\ge3}
\mu^{(D-1)(\frac n2 -1)}\,v_nq^n
$$
and Fermi-level ${\cal E}$ is determined by the equation
$$
\int\frac{dp\,dq}{2\pi}\,\theta\bigl({\cal E}-H(p,q)\bigr)\,=1.
$$
Note that $E=0$ for $M=0$ and $V(q)=0$.

\noindent
{\bf 5.} Consider now the two-point correlator of the scalar fields:
\begin{eqnarray}
\label{Kxi}
K^{(D)}(x)\,&=&\,\frac 1Z\int \prod_{\mu=0}^{D-1} dA_\mu \,c_\mu\,d\varphi\,
Tr\bigl[\varphi(x)\varphi(0)\bigr]\,\delta\bigl(A_0\bigr)\\
&\times&\,
\exp \left(\frac{2\pi}{\Lambda}\right)^D\,
Tr\left\{\frac 12\sum_\mu \bigl[D_{\mu},\,\varphi\bigr]^2
-\frac 12 M^2\,\varphi^2-V(\varphi)\right\}.\nonumber
\end{eqnarray}
The variables changing (\ref{change}) brings the integral to the form
\begin{eqnarray}
\label{Kxf}
K^{(D)}(x)\,&=&\,\frac 1Z\int dV_0\,
\prod_{\mu=1}^{D-1} d\widetilde{V}_\mu \,dU\,d\varphi_d
\Delta^2\bigl(\varphi_d\bigr)
\delta \left[\frac{N_c}g
\left(P_0\,-\,V_0^{-1}P_0 V_0 \right)\right] \nonumber\\
&\times&\,Tr\bigl[e^{iPx}U^{-1}\varphi_d U e^{-iPx}\cdot U^{-1}\varphi_d U
\bigr]\\
&\times&\,\exp \left(\frac{2\pi}{\Lambda}\right)^D\,Tr \left\{\frac 12
\bigl[P_0,\,V_0 U^{-1}\varphi U V_0^{-1}\bigr]^2\right.\nonumber\\
&&+\,\left.\frac 12\sum_{\mu=1}^{D-1}\bigl[P_\mu\,,\widetilde{V}_{\mu}\varphi_d
\widetilde{V}_{\mu}^{-1}]^2\,-\,\frac 12 M^2\,\varphi^2-V(\varphi)\right\}.
\nonumber
\end{eqnarray}
In contrast to the partition function case the pre-exponential factor
prevents to factorize the gauge-fixing term with the integral over $V_0$.
Nevertheless the "dimensional scaling property" (\ref{scale}) can be used
again to reduce the product of the independent integrals $Z_{\varphi}^{D-1}$
to the single one:
\begin{eqnarray}
K^{(D)}(x)\,&=&\,\frac 1Z\int dV_0\,d\widetilde{V}_1 \,dU\,d\varphi_d
\Delta^2\bigl(\varphi_d\bigr)
\delta \left[\frac{N_c}g
\left(P_0\,-\,V_0^{-1}P_0 V_0 \right)\right]\nonumber \\
&\times&\,Tr\bigl[e^{iPx}U^{-1}\varphi_d U e^{-iPx}\cdot U^{-1}\varphi_d U
\bigr]\nonumber\\
&\times&\,\exp \left(\frac{2\pi}{\Lambda}\right)^D\,Tr \left\{\frac 12
\bigl[P_0,\,V_0 U^{-1}\varphi U V_0^{-1}\bigr]^2\right.\nonumber\\
&&+\,\left.\frac 12(D-1)\bigl[P_1\,,\widetilde{V}_1\varphi_d
\widetilde{V}_1^{-1}]^2\,-\,\frac 12 M^2\,\varphi^2-V(\varphi)\right\}.
\nonumber
\end{eqnarray}
Now the steps leading from (\ref{Kxi}) to (\ref{Kxf}) can be repeated
in the inverse order which yeilds
\begin{eqnarray}
K^{(D)}(x)\,&=&\,\frac 1Z\int \prod_{\mu=0}^1 dA_\mu \,c_\mu\,d\varphi\,
Tr\bigl[\varphi(x)\varphi(0)\bigr]\,\delta\bigl(A_0\bigr)\nonumber\\
&\times&\,
\exp \left(\frac{2\pi}{\Lambda}\right)^D\,
Tr\left\{\frac 12\bigl[D_0,\,\varphi\bigr]^2+
\frac 12(D-1)\,\bigl[D_1,\,\varphi\bigr]^2\right.\nonumber\\
&-&\,\left.\frac 12 M^2\,\varphi^2-V(\varphi)\right\}.\nonumber
\end{eqnarray}
After rescaling
\begin{eqnarray}
\label{rescal1}
P_0\,&\to&\,\sqrt{D-1}\,P_0 \\
\label{rescal2}
\varphi\,&\to&\frac 1{\sqrt{D-1}}\,\mu^{\frac D2 -1}\varphi
\end{eqnarray}
the integral takes a more symmetric form \footnote{The first line here is the
quenched momentum analog of $x_0 \to x_0 \sqrt{D-1}$ replacement. Although
the integration region is no more a hypercube in the momentum space it
is not essential for a ultraviolet convergent $2D$ theory.}
\begin{eqnarray}
K^{(D)}(x)\,&=&\,\mu^{D-2}\frac 1Z\,\int \prod_{\mu=0}^1 dA_\mu \,c_\mu\,
d\varphi\,Tr\bigl[\varphi(\sqrt{D-1}\,x_0,x)\varphi(0)\bigr]\,
\delta\bigl(A_0\bigr)\nonumber\\&\times&\,
\exp \left(\frac{2\pi}{\Lambda}\right)^D\,
Tr\left\{\frac 12\sum_{\mu=0}^1\bigl[D_\mu,\,\varphi\bigr]^2
-\,\frac 12 \frac{M^2}{D-1}\,\varphi^2-
\widetilde{V}(\varphi)\right\}\nonumber
\end{eqnarray}
where
\begin{equation}
\label{V}
\widetilde{V}(\varphi)\,=\,\sum_{n\ge3}\bigl(D-1\bigr)^{-\frac n2}\,
\mu^{\bigl(\frac D2-1\bigr)(n-2)}\,v_n\varphi^n
\end{equation}
or
$$
K^{(D)}(x_0,x_1,0,\ldots,0)\,=\,\mu^{D-2}\frac 1{D-1}\,K^{(2)}(\sqrt{D-1}x_1,x_2).
$$
Here $K^{(2)}$ is the correlator in two-dimensional theory.

Thus the two-point scalar correlator in the $D$-dimensional induced QCD
coincides with the same (up to rescaling) correlator calculated
in the two-dimensional induced QCD. This result clearly holds for any
$n$-point scalar correlator provided all the points lay in the same
two-dimensional plane.

\noindent
{\bf 6.} Unfortunately the above methods based on the "scaling property"
(\ref{scale})
are insufficient for gluon correlators. The tensor structure of the vector
field correlators in D-dimensional space is much more rich than for $D=2$
and they can not be reduced to a two-dimensional theory. However there is a
special kinematic to which a little modification of the property
(\ref{scale}) can be applied.

Consider again a two-point correlator
\begin{equation}
\label{dD}
d^{\,(D)}(x)\,=\,\langle\,Tr\,n_\mu A_\mu(x)\,n_\nu A_\nu(0)\,\rangle
\end{equation}
where $n$ is an arbitrary vector of the form $n_\mu = (0,n_1,\ldots,n_{D-1})$
(recall the gauge is $A_0 = 0$) normalized as $n_\mu^2 = (D-1)$. One can
always choose the coordinate axes in such a way that $n_\mu = (0,1,
\ldots,1)$. The vector $x$ in (\ref{dD}) will be assumed to be of the form
$x=(x_0,x_1,0,\ldots,0)$ in this basis, so $(xn)^2=x_1=x_\perp^2$.

It is convenient to deal with the generating functional which in the quenched
momentum prescription is
\begin{eqnarray}
\label{ZJ}
Z_J\,&=&\,\int \prod_{\mu=0}^{D-1} dA_\mu \,c_\mu\,\delta(A_0)\,d\varphi\,
\\ &\times&\,
\exp \left(\frac{2\pi}{\Lambda}\right)^D\,
Tr\left\{\frac 12\sum_\mu\bigl[D_\mu,\,\varphi\bigr]^2-
\frac 12 M^2\,\varphi^2-V(\varphi)\right.\nonumber \\
&&+\,\left.\mu^D J_2\,\sum_\mu A_\mu(x)\,+\,\mu^D J_1\,\sum_\mu A_\mu(0)
\right\}.\nonumber
\end{eqnarray}
The correlator is clearly given by differentiating (\ref{ZJ}) with respect
to the matrixes $J_{1,2}$:
$$
d^{\,(D)}(x)\,=\,Tr\,\left.\frac{\delta}{\delta J_2}\frac{\delta}{\delta J_1}\,
\cdot \,Z_J\right|_{J_{1,2}=0}.
$$
After variables shifting and resolving the $c_\mu$ constraints $Z_J$ takes
the form
\begin{eqnarray}
\label{Jproduct}
Z_J\,&=&\,\exp\left\{\frac{\sqrt{N_c}}g\,Tr\bigl(J_2+J_1\bigr)\,
\sum_\mu P_\mu\,\right\}\\
&\times&\exp \left(\frac{2\pi}{\Lambda}\right)^D\,
Tr\left\{\frac 12\bigl[P_0,\,\varphi\bigr]^2\,
-\,\frac 12 M^2\,\varphi^2\,-\,V(\varphi)\right\}\nonumber\\
&\times&\int \prod_{\mu=0}^{D-1} dV_\mu\,\exp \left(\frac{2\pi}
{\Lambda}\right)^D\,
\left\{ Tr\frac 12 \bigl[V^{-1}_\mu P_\mu V_\mu\,,\varphi\bigr]^2\,\right.
\nonumber \\
&&-\,\left. \mu^D \frac{\sqrt{N_c}}g\,J_2\,e^{iPx}V^{-1}_\mu P_\mu V_\mu
e^{-iPx}\,-\,\mu^D \frac{\sqrt{N_c}}g\,J_1\,V^{-1}_\mu P_\mu V_\mu \right\}.
\nonumber
\end{eqnarray}
The internal integral here is again the product of $(D-1)$ equal "angular"
integrals. Although they differ from $Z_\varphi$ (\ref{zphi}) by additional
terms in the exponent these terms are of the order $N_c^2$ ($J_{ik}\sim 1$)
therefore the same type saddle point behavior is natural to be assumed for
these integrals too. The continuous representation like (\ref{zcont}) can
be also written for each of them as
$$
\int{\cal D}V\exp\,N_c^2\,\left(\frac{2\pi}{\Lambda}\right)^D\,
\int d\Omega \left[\frac 12\bigl\{\,p\,,\,V(\varphi)\,\bigr\}^2
+\mu^D J_2(x)V(p) + \mu^D J_1V(p) \right]
$$
where $V(p)$ denotes as before the action of the area preserving
diffeomorphism $V$ on the continuous image (function on the unit sphere)
of the matrix $P_\mu$, $J_2(x)$ and $J_1$ are the continuous images of the
matrixes $e^{-iPx} J_2 e^{iPx}$ and $J_1$ respectively. This formula enables
to replace the product of the integrals in (\ref{Jproduct}) by the single
integral for the momentum $P_1$. Introducing then the field $A_1$ through
the relation
$$
V^{-1} P_1 V\,=\,\frac 1i D_1\,=\,P_1\,+\,\frac g{\sqrt{N_c}}\,
\mu^{\frac D2-1}\,A_1
$$
and making rescaling (\ref{rescal1}), (\ref{rescal2}) the generating
functional takes a final form
\begin{eqnarray}
Z_J\,&=&\,N \exp\left\{\frac{\sqrt{N_c}}g\,Tr\bigl(J_2+J_1\bigr)\,
\sum_\mu P_\mu\,-\,(D-1)P_1\right\}\nonumber\\
&\times&\int \prod_{\mu=0}^1 dA_\mu \,c_\mu\,\delta(A_0)\,d\varphi\,
\exp \left(\frac{2\pi}{\Lambda}\right)^2\,
Tr\left\{\frac 12\sum_\mu\bigl[D_\mu,\,\varphi\bigr]^2-
\frac 12 \frac{M^2}{D-1}\,\varphi^2\right.\nonumber \\
&&-\widetilde{V}(\varphi)\,\left.(D-1)\mu^{\frac D2+1}\,J_2 A_1(x)\,
-\,(D-1)\mu^{\frac D2+1} J_1 A_1 \right\}\nonumber
\end{eqnarray}
where $\widetilde{V}(\varphi)$ is given by (\ref{V}) and the factor
$N$ does not
depend on $J_{1,2}$. It gives for the correlator
\begin{equation}
\label{dDd2}
d^{\,(D)}(x_0,x_1,0,\ldots,0)\,=\,(D-1)^2\mu^{D-2}\,\langle Tr A_1(\sqrt{D-1}
\,x_0,x_1)\,A_1(0)\rangle_2
\end{equation}
where the right hand side means the correlator in the two-dimensional induced
QCD.

Note that a two-point gluon correlator in the dimension $D$
$$
d^{\,(D)}_{\mu\nu}(x)\,=\,\langle Tr A_\mu(x)\,A_\nu(0)\rangle
$$
is determined by two scalar functions
$$
d^{\,(D)}_{\mu\nu}(x)\,=\,x^{\perp}_\mu x^{\perp}_\nu\, d^{\,(D)}_2(x_0,
x_\perp^2)\,+\,\delta^\perp_{\mu\nu}\,d^{\,(D)}_0(x_0,x_\perp^2).
$$
The formula (\ref{dDd2}) implies one relation between them:
$$
d^{\,(D)}_2(x_0,x_\perp^2)\,+\,d^{\,(D)}_0(x_0,x_\perp^2)\,=
$$
$$
=\,(D-1)^2\mu^{D-2}\,\bigl[d^{\,(2)}_2(\sqrt{D-1}\,x_0,x_\perp^2)+
d^{\,(2)}_0(\sqrt{D-1}\,x_0,x_\perp^2)\bigr].
$$

There is another way to derive this relation. Instead of (\ref{dD}) one can
start from the correlator
$$
d^{\,(D)}_{\mu\mu}(x)\,=\,
\sum_\mu \langle Tr A_\mu(x)\,A_\mu(0)\rangle\,=
\,\frac{\sqrt{N_c}}g
\langle\,Tr\,\bigl[P_\mu P_\nu\,- \, D_\mu(x)D_\mu(0)\bigr]\,\rangle
$$
and introduce the generating functional
$$
Z_\sigma\,=\,\langle\exp\left\{\sigma\,Tr D_\mu(x)D_\mu(0)\right\}\rangle,
$$
$$
d^{\,(D)}_{\mu\mu}(x)\,=\,\left.\frac{\partial}{\partial\sigma}\,Z_\sigma
\right|_{\sigma = 0}.
$$
The integral for $Z_\sigma$ allows for the dimensional reduction like
(\ref{ZJ}) since the "angular" part of it is again the product of $(D-1)$
independent equal integrals.

One can combine both these trick to derive similar relations for higher
correlators with the gluon fields. However there is a lot of tensor
structures which are not involved in such relations at all. For instance,
the triple gluon vertex does not contribute to any of them because of
antisymmetry over the index permutations.

\noindent
{\bf 7.} Essential reduction of the space degrees of freedom occurring in the
induced QCD is the main result of this paper. All D-dimensional induced QCD
theories turn out to be related through the equality (\ref{scale}).
The reason for the "scaling property" (\ref{scale}) lies, probably, in the
fact that coordinate and momentum operators can be approximated by large
$N_c$ matrixes. Although there are no finite order representation for the
Heizenberg algebra one can construct the matrixes $X_\mu$ and $P_\nu$ from
$SU(N_c)$ algebra for which $[X_\mu,P_\nu] = i\delta_{\mu\nu} + O(1/N_c)$
\cite{eguchi}. It explains why space disappears in the induced QCD: the
space-time transformation can be mapped for any dimension D into the same
$SU(N_c)$ group. The quenched momentum prescription is a particular choice
of such a map. One should stress the point however that the dimensional
scaling is valid only for the gauge theories because only for them the
"angular" (over unitary matrixes) integrals are carried out separately
for each field component. It would be impossible for a scalar theory
where the unitary rotation is common for all space directions. It is
the additional unitary integrals in the QCD that "absorb" the space degrees
of freedom.

Thus the induced QCD results into the low dimensional theory - one
dimensional quantum mechanics for the partition function case and two
dimensional theory for the scalar correlators. The latter fact is of
special interest because the two dimensional QCD models with a scalar
matter in adjoint representation have received a recent attention
due to a similarity they have with 4D gauge theory \cite{QCD_2,QCD_2n}.
The scalar
degrees of freedom "compensate" the absence of transverse gluons in
2D models. From the other hand it is a commonplace that the linear
potential resulting from the Coulomb interaction in 2D theories has no
relation to 4D confinement. The dimensional reduction hints that such a
relation might exist. Indeed, the interaction potential between the large
mass matter particles can be extracted from the amplitude of their elastic
$2\to2$ scattering. It is the amplitude that allows for
the two-dimensional reduction in the induced QCD. The Coulomb force
originates from $G_{\mu\nu}^2$ term in the
effective lagrangian which appears after integration over the scalar field.
However there is not the logarithmic divergency when D = 2, and the first
term does not dominate in the loop expansion in which the higher terms are
related to the presence of the non-Coulomb degrees of freedom.

The natural question here is in what extent these results are valid for
the real, not induced, QCD. Unfortunately the naive picture supposing all
the terms except the first one in the loop expansion to be suppressed by
the large scalar mass $M$ is inconsistent. The scalar loops result into the
point-like gluon interaction only if the gluons' momenta are much smaller
than the $M$. Therefore the induced QCD has to imply two typical scales to
be equivalent to the gluodynamics -- the ultraviolet cutoff $\Lambda$ for
the scalar particles and their mass $M\ll\Lambda$ for the gluons. However
the gluons' momenta circulating in the internal loops are restricted
really by the $\Lambda$ rather than the $M$ value. That is why the
expressions (\ref{fe}), (\ref{feh}) for the vacuum energy density can not be
immediately applied to the QCD since the hard gluon loops contribution is
not separated from them.

A possible way to connect the induced QCD with the real gluodynamics is
to adopt a slightly different approach. One can start from the scalar QCD
treating the scalars on an equal footing with the gluons so the lagrangian
will encounter the scalars selfinteraction $h\varphi^4$. The constant $h$
as well as the gluon coupling $g$ are the bare constants which have to be
taken to be functions of the regularization parameter $\Lambda$. They
values are determined by regularization procedure: the gluon and scalar
dressed vertexes and the residues of the propagators poles are fixed at
some external momentum scale $\mu$ referred to as a normalization point.
The position of the scalar propagator pole is fixed too as a physical
scalar mass. If the bare constants are adjusted to keep all the fixed at the
$\mu$ point values when the ultraviolet cutoff $\Lambda\to\infty$, then
according to the renormalization theory the Green functions will be finite.
It is the dressed vertexes values at the normalization point rather than the
bare constants that play the physical charges role in the theory.

The bare gluon coupling is zero in the induced QCD, therefore the total
number of the bare parameters is insufficient to satisfy all the
normalization conditions. However one can lift one of them, say, not to fix
the position of the scalar propagator poles using the bare scalar mass as a
parameter to keep the gluon vertex value. Suppose the renormalization scheme
can be carried out for all the vertexes and residues while the physical
scalar mass turns out to be of the order of $\Lambda$. In this case the
renormalized theory will be surely equivalent to the QCD. Even if the more
complicated situation occurs it will be of interest to study what is the
limiting theory. The obtained here results enable to investigate this problem
in the framework of the two-dimensional scalar QCD.

There are several difficulties however to proceed in this manner. It is
unclear if the triple gluon vertex allows to be reduced to the
two-dimensional theory like the propagators do. The possible way to
circumvent this obstacle is to determine the gluon physical coupling through
the interaction with a large massive matter field, that is through the
Coulomb potential. An other probably more serious difficulty is the absence
of an exact solution of the massive two-dimensional QCD although there are
numerical investigations of it \cite{QCD_2n}.

Note in the conclusion that the induced QCD could be interesting in itself
as a toy model even without the direct equivalence with the gluodynamics.
Indeed it provides an example of the four-dimensional theory asymptotically
free at the small distances (in some sense the asymptotic freedom in the
induced QCD is even more strong than in the gluodynamics \cite{asfr}) which
exhibits a confinement-like behavior at the large ones.

\end{document}